# Structural/microstructural, optical and electrical investigations of Sb-SnO$_2$ thin films deposited by spray pyrolysis

Sushant Gupta[1], B.C. Yadav*[1], Prabhat K. Dwivedi[2] and B. Das[3]

[1]*Department of Applied Physics, School for Physical Sciences,*
*Babasaheb Bhimrao Ambedkar Central University, Lucknow-226025, U.P., India*

[2]*DST Unit on Nanosciences, Department of Chemical Engineering,*
*Indian Institute of Technology Kanpur, Kanpur-208016, U.P., India*

[3]*Department of Physics, University of Lucknow,*
*Lucknow-226007, U.P., India*

*\*Email:balchandra_yadav@rediffmail.com*
*Phone: +91-522-2998125, Mobile: +919450094590*

## Abstract

The structural, optical and electrical properties of spray deposited antimony (Sb) doped tin oxide (SnO$_2$) thin films, prepared from SnCl$_4$ precursor, have been studied as a function of antimony doping concentration. The doping concentration was varied from 0 to 1.5 wt.% of Sb. The analysis of x-ray diffraction patterns revealed that the as deposited doped and undoped tin oxide thin films are pure crystalline tetragonal rutile phase of tin oxide which belongs to the space group P4$_2$/mnm (number 136). The surface morphological examination with field emission scanning electron microscopy (FESEM) revealed the fact that the grains are closely packed and pores/voids between the grains are very few. The transmittance spectra for as-deposited films were recorded in the wavelength range of 200 to 1000 nm. The transmittance of the films was observed to increase from 57% to 68% (at 800 nm) on initial addition of Sb (up to [Sb]/[Sn] = 0.5 wt.%) and then it is decreased for higher level of antimony doping ([Sb]/[Sn] > 0.5 wt.%). The sheet resistance of tin oxide films was found to decrease from 48 Ω/sq for undoped films to 8 Ω/sq for antimony doped films.

## 1. Introduction

Transparent conducting oxides (TCOs) are solid-state oxides that combine low electrical resistance with high optical transparency in the visible range of the electromagnetic spectrum [1]. These properties are sought in a number of applications; notably as electrode materials in solar cells, light emitting diodes, flat panel displays, and other optoelectronic devices where an electric contact needs to be made without obstructing photons from either entering or escaping the optical active area and in transparent electronics such as transparent field effect transistors [2-12]. Another property of TCOs is that although they are transparent in the visible light they are highly reflective for infrared light. This property is responsible for today's dominant use of TCO as an energy conserving material. TCO coated architectural windows, for instance, allow the light to transmit but keeping the heat out or in the building depending on the climate region. More sophisticated architectural windows, so-called smart windows, rely on TCOs to electrically contact electrochromic films that are changing their coloring and transparency by applying a voltage across the films [13-15].

There is a large number of TCOs, the most commonly known ones are the binary systems, i.e. $SnO_2$, $ZnO$, $In_2O_3$, $Ga_2O_3$, and $CdO$ [16,17]. A large variety of ternary ($Cd_2SnO_4$, $CdSnO_3$, $CdIn_2O_4$, $Zn_2SnO_4$) and more complex TCO materials are being developed [18-21] and continuous efforts are being made to find p-type conducting TCOs [22] in addition to the above-mentioned n-type materials. For different applications different materials may possess advantageous properties [23]. Hartnagel et al. gave a review of properties and preparation procedures for TCOs [1]. For practical use as transparent electrodes in devices such as solar cells, flat panel displays, and light emitting diodes, a TCO must have a resistivity of less than $10^{-3}$ Ω cm and over 80% transmittance in the visible range [18]. Indium tin oxide (ITO) is the current industrial standard material for transparent electrodes as thin films can be produced with resistivities of the order of $10^{-5}$ Ω cm. However, due to the expense and scarcity of indium, alternatives need to be found. Among the available TCOs, $SnO_2$ seems to be more appropriate because they are quite stable toward atmospheric conditions, chemically inert, mechanically hard and can resist high temperature but its conductivity does not yet approach to that of ITO [24].

Thin films of $SnO_2$ can be prepared by many techniques, such as chemical vapor deposition [25], sputtering [26], sol-gel [27], reactive evaporation [28], pulsed laser ablation [29], screen printing technique [30], and spray pyrolysis [31]. Among these, spray pyrolysis is the most convenient method because of its simplicity, low cost, easy to add doping materials, and the possibility of varying the film properties by changing composition of starting solution. Otherwise, this method is promising for high rate and mass production capability of uniform large area coatings in industry.

The main objective of this work is to prepare high conducting Sb doped $SnO_2$ thin films by chemical spray pyrolysis method with different doping levels of Sb ([Sb]/[Sn] =

0.0 to 1.5 wt.%) and explore its structural, morphological, electrical and optical properties.

**2. Experimental details**

Thin films of pure and Sb doped $SnO_2$ were deposited by spray pyrolysis method. The quality of these films depends on various process parameters such as spray rate, substrate temperature and the ratio of the various constituents in the solutions. Since the deviation from stoichiometry due to oxygen vacancies [32-34] makes tin oxide thin films to possess semiconducting nature, it is very essential that the complete oxidation of the metal should be avoided in order to obtain films with good conductivity. This is generally achieved by adding appropriate reducing agents. Methanol was used as the reducing agent in the present work.

The substrate temperature also plays an important role in the film formation. When the substrate temperature is below $350^oC$, the spray falling on the substrate will undergo incomplete thermal decomposition (oxidation) giving rise to a foggy film whose transparency as well as electrical conductivity will be very poor. If the substrate temperature is too high ($> 500^oC$) the spray gets vaporized before reaching the substrate and the film becomes almost powdery. Whereas at substrate temperature in the range of $350-500^oC$ the spray reaches the substrate surface in the semi vapour state and complete oxidation will take place to give clear $SnO_2$ film as a final product. Keeping these facts in mind, we optimize substrate temperature at $425^oC$.

An amount of 17.529 gm of $SnCl_4.5H_2O$ (Merck purity > 98 %) was dissolved in 5 ml of concentrated hydrochloric acid (Merck, min 35% GR) by heating at $90^oC$ for 15 minutes. The addition of HCl rendered the solution transparent, mostly, due to the breakdown of the intermediate polymer molecules [35]. The transparent solution thus obtained and subsequently diluted by methanol, served as the precursor. To achieve Sb doping, antimony trichloride ($SbCl_3$) was dissolved in isopropyl alcohol and added to the precursor solution. The amount of ($SbCl_3$) to be added depends on the desired doping concentration. The doping concentration was varied from 0-1.5 wt.%. The amount of spray solution was made together 50 ml. For each concentration the reproducibility of the films were verified by repeating the experiments several times. Microscope glass slides (2.0×2.5 $cm^2$), cleaned with organic solvents, were used as substrates. During deposition, the solution flow rate was maintained at 0.2 ml/min by the nebulizer (particle size 0.5 to 10 μm). The distance between the spray nozzle and the substrate as well as the spray time was maintained at 3.0 cm and 15 minutes respectively. The thickness of the films was observed to be at the range of 1 − 10 μm.

Sb doped $SnO_2$ films have an appearance of bluish coloration due to the addition of Sb atoms which was predictable since it was reported by many authors [36, 37].

The gross structure and phase purity of pure and Sb doped $SnO_2$ films were examined by glancing angle x-ray diffraction (GAXRD) technique using a Philips x-ray diffractometer (X' Pert PRO, Model PW 3040). In the present study, all the XRD patterns of undoped and Sb doped $SnO_2$ thin films were recorded in the 2θ range from $20^o$ to $60^o$ with a $1^o$ glancing angle, this angle was kept small ($1^o$) to ensure that maximum signal comes from films rather than from the substrates. The experimental peak positions were compared with the data from the database Joint Committee on Powder Diffraction Standards (JCPDS) and Miller indices were assigned to these peaks. Morphologies of as-deposited films were investigated by field emission scanning electron microscopy (ZEISS-FESEM). The transmission and absorption spectra of pure and Sb doped $SnO_2$ films were recorded using Dual beam UV-Vis spectrometer (Cary 50) in the wavelength ranging from 200 to 1000 nm. Room temperature sheet resistance was determined by four probe method employing van der Pauw geometry.

4. **Results and discussion**

Figure 1 shows the x-ray diffraction patterns of pure $SnO_2$ and Sb-$SnO_2$ with various Sb concentration (0.5 – 1.5 wt.%). The analysis of x-ray diffraction patterns revealed that the as deposited doped and undoped tin oxides films are pure crystalline tetragonal rutile phase of tin oxide (JCPDS card no. 041-1445) which belongs to the space group $P4_2/mnm$ (number 136). No obvious reflection peaks from impurities, such as unreacted Sn, Sb or other oxide phases such as $Sb_2O_5$ or $Sb_2O_3$ are detected, indicating high purity of the product. It is perceptible from the XRD patterns of Figure 1 that the undoped as well as doped tin oxide films grow along the preferred orientation of (110). The presence of other orientations such as (101), (200) and (211) have also been detected with considerable intensities for both doped and undoped tin oxide films.

We have calculated the lattice parameters using XRD peaks such as (110), (101), (200) and (211) shown in Figure 1. The calculated lattice parameters of Sb-$SnO_2$ ([Sb]/[Sn] = 0.0, 0.5, 1.0, 1.5 wt.%) are shown in Table 1. A small decrease in the lattice parameters of the tetragonal unit cell has been observed with increasing Sb content (Figures 2 & 3). For example for [Sb]/[Sn] = 0.0 wt.%, a = 4.7384 Å, c = 3.1899 Å whereas for [Sb]/[Sn] = 1.5 wt.%, a = 4.7295 Å, c = 3.1803 Å. This may possibly occur due to the difference in ionic radii of $Sn^{4+}$ (0.72 Å) and $Sb^{5+}$ (0.62 Å) ions.

The diffraction peaks are markedly broadened, which indicates that the crystalline sizes of deposited films are small. Crystallite size was automatically calculated from x-ray diffraction data using the Debye-Scherrer formula, [38]:

$$D_{hkl} = 0.9\lambda/\beta \cos\theta, \qquad (1)$$

where $\lambda$ is the x-ray wavelength (1.5418 Å for $CuK_\alpha$), $\theta$ is the Bragg angle and $\beta$ is the full width of the diffraction line at half its maximum intensity (FWHM). The average crystallite size of [Sb]/[Sn] = 0.0, 0.5, 1.0 and 1.5 wt.% were calculated using Eq.1 as 40, 35, 25 and 20 nm respectively.

The FESEM images of undoped and Sb doped $SnO_2$ thin films deposited by chemical spray pyrolysis technique at substrate temperature of 425 °C are shown in Figures 4 to 7. These FESEM images reveal that the grains are closely packed and pores/voids between the grains are very few. The particle size shown by FESEM was higher as compared with that calculated from the XRD results. This was because of the fact that the XRD gave the average mean crystallite size while FESEM showed agglomeration of the particles. The XRD and FESEM data can be reconciled by the fact that smaller primary particles have a large surface free energy and would, therefore, tend to agglomerate faster and grow into larger grains.

The transmittance (T) spectra of Sb-$SnO_2$ ([Sb]/[Sn] = 0.0, 0.5, 1.0, 1.5 wt.%) thin films as a function of wavelength ranging from 200 to 1000 nm is shown in Figure 8. The transmittance value of 56.96% (at 800 nm) for the pure tin oxide films is found to increase to 68.10% (at 800 nm) on the addition of 0.5 wt.% of antimony. But the transmittance is found to decrease gradually if the antimony concentration is increased above 0.5 wt.%. The transmittance for selected wavelengths (600, 700, 800, and 900 nm) is plotted in Figure 9 as a function of antimony doping concentration. It is evident from the figure that for all the wavelengths, the transmittance of the pure tin oxide films is increased for initial doping (0.5 wt.%) of antimony whereas the transmittance decreases for the next higher levels of doping. The effect of antimony doping, with [Sb]/[Sn] > 0.5wt.%, on decreasing the transmittance continuously within our range of doping, with intense blue in color must result from light absorption in the film. It was reported by Kojima et al. [37] that when a material contains an element in two different oxidation states or in a mixed oxidation state, it manifests abnormally deep and intense coloration. The reason is that electron transfer between the different oxidation states of the element, namely $Sb^{5+}$ and $Sb^{3+}$ in our case, causes intense light absorption. Hence the decrease in the transmittance of Sb-$SnO_2$ films with increase in doping concentration (in the present study) may be due to the increasing absorption.

The variation of the optical absorption coefficient α with photon energy hν was obtained using the absorbance data for various films. The absorption coefficient α may be written as a function of the incident photon energy hν [39]:

$$\alpha = [A(h\nu - E_g)^n]/h\nu \qquad (2)$$

where A is a constant which is different for different transitions indicated by different values of n, and $E_g$ is the corresponding bandgap. For direct transitions n = ½ or n = 2/3, while for indirect ones n = 2 or 3, depending on whether they are allowed or forbidden, respectively [39]. Many groups have used the above formula to calculate the bandgap of $SnO_2$ films and reported that $SnO_2$ is a direct bandgap material [40-44]. The bandgap can be deduced from a plot of $(\alpha h\nu)^2$ versus photon energy (hν). Better linearity of these plots suggests that the films have direct band transition. The extrapolation of the linear portion of the $(\alpha h\nu)^2$ vs. $h\nu$ plot to $\alpha = 0$ will give the bandgap value of the films [45].

Figures 10(a) to 10(d) shows the $(\alpha h\nu)^2$ versus photon energy ($h\nu$) plot for pure and antimony (Sb) doped $SnO_2$ films. The linear fits obtained for these plots are also depicted in the figures. The bandgap ($E_g$) values for the Sb-$SnO_2$ films with antimony (Sb) concentrations [Sb]/[Sn] = 0.0, 0.5, 1.0 and 1.5 wt.% are 4.119, 4.137, 4.192 and 4.283 eV respectively. From Figures 10(a) to 10(d), it is observed that the bandgap of the films increases with the increase in concentration of antimony (Sb). This is in agreement with the Burstein-Möss effect [46, 47]. The bandgap is plotted as a function of increasing Sb concentration in Figure 11, and also the $E_g$ values are given in Table 2.

The sheet resistance value measured for the Sb-$SnO_2$ films with antimony (Sb) concentrations [Sb]/[Sn] = 0.0, 0.5, 1.0 and 1.5 wt.% are 48, 23, 16 and 8 $\Omega$/sq. respectively, this variation is plotted in Figure 12. From Figure 12, it is observed that the sheet resistance of the films decreases with the increase in concentration of antimony (Sb). The possible mechanism of this variation can be explained as, when $SnO_2$ is doped with Sb a part of the lattice $Sn^{4+}$ atoms are replaced by $Sb^{5+}$ resulting in the generation of conduction electrons and thus reducing the sheet resistance [33, 37].

**4. Conclusions**

Thin films of pure and antimony doped tin oxide were prepared by spray pyrolysis technique from $SnCl_4$ precursor. The analysis of x-ray diffraction patterns revealed that the as deposited doped and undoped tin oxide thin films are pure crystalline tetragonal rutile phase of tin oxide (JCPDS card no. 041-1445) which belongs to the space group $P4_2/mnm$ (number 136). A small decrease in the lattice parameters of the tetragonal unit cell has been observed with increasing Sb content. This possibly occurs due to the difference in ionic radii of $Sn^{4+}$ (0.72 Å) and $Sb^{5+}$ (0.62 Å) ions. The average crystallite size of [Sb]/[Sn] = 0.0, 0.5, 1.0 and 1.5 wt.% were calculated as 40, 35, 25 and 20 nm respectively. Surface morphology examination with FESEM in scanning mode revealed the fact that the grains are closely packed and pores among them are very few. The transmittance increases initially with the increase in doping concentration and then decreases for higher doping levels which is attributed to light absorption. The energy bandgap of Sb doped $SnO_2$ films were calculated from optical absorption spectra by UV-Vis absorption spectroscopy. Upon increasing the Sb concentration, the bandgap of the films was found to increase from 4.119 eV to 4.283 eV. The sheet resistance of tin oxide films was found to decrease from 48 $\Omega$/sq for undoped films to 8 $\Omega$/sq for antimony doped films.

**Acknowledgements**

Authors gratefully acknowledge to Prof. R.G. Sonakawade, Coordinator, Department of Applied Physics, Babasaheb Bhimrao Ambedkar Central University, Lucknow for their encouragement and DST Unit on Nanosciences, Department of Chemical Engineering, Indian Institute of Technology Kanpur for providing the facilities.

**Figure captions:-**

Figure 1: X-ray diffraction pattern for Sb-doped $SnO_2$ films for different concentrations of dopant.

Figure 2: The variation of lattice parameter 'a' (or 'b') versus Sb - concentration shows minute decrement in lattice parameter 'a' (or 'b').

Figure 3: The variation of lattice parameter 'c' versus Sb – concentration shows minute decrement in lattice parameter 'c'.

Figure 4: FESEM image of pure $SnO_2$ thin film.

Figure 5: FESEM image of Sb doped $SnO_2$ thin film with [Sb]/[Sn] = 0.5 wt.%.

Figure 6: FESEM image of Sb doped $SnO_2$ thin film with [Sb]/[Sn] = 1.0 wt.%.

Figure 7: FESEM image of Sb doped $SnO_2$ thin film with [Sb]/[Sn] = 1.5 wt.%.

Figure 8: Transmittance spectra of $SnO_2$: Sb ([Sb]/[Sn] = 0.0, 0.5, 1.0, 1.5 wt.%) thin films as a function of wavelength.

Figure 9: Transmittance for selected wavelengths of Sb-$SnO_2$ thin films as a function of Sb doping.

Figure 10: $(\alpha h\nu)^2 [eV^2]$ versus photon energy (hν) [eV] curve for the (a) [Sb]/[Sn] = 0.0 wt.%, (b) [Sb]/[Sn] = 0.5 wt.%, (c) [Sb]/[Sn] = 1.0 wt.%, (d) [Sb]/[Sn] = 1.5 wt.%. The direct energy bandgap $E_g$ is obtained from the extrapolation to α = 0.

Figure 11: Variation of bandgap ($E_g$) as a function of increasing antimony (Sb) concentration. The bandgap of the films increased from 4.119 to 4.283 eV due to Sb doping.

Figure 12: Sheet resistance of Sb-$SnO_2$ thin films as a function of Sb doping.

**Table Captions:-**

Table-1. Lattice parameters of Sb: $SnO_2$ ([Sb]/[Sn] = 0.0, 0.5, 1.0, 1.5 wt.%).

Table-2. Optical bandgap of the Sb doped $SnO_2$ films.

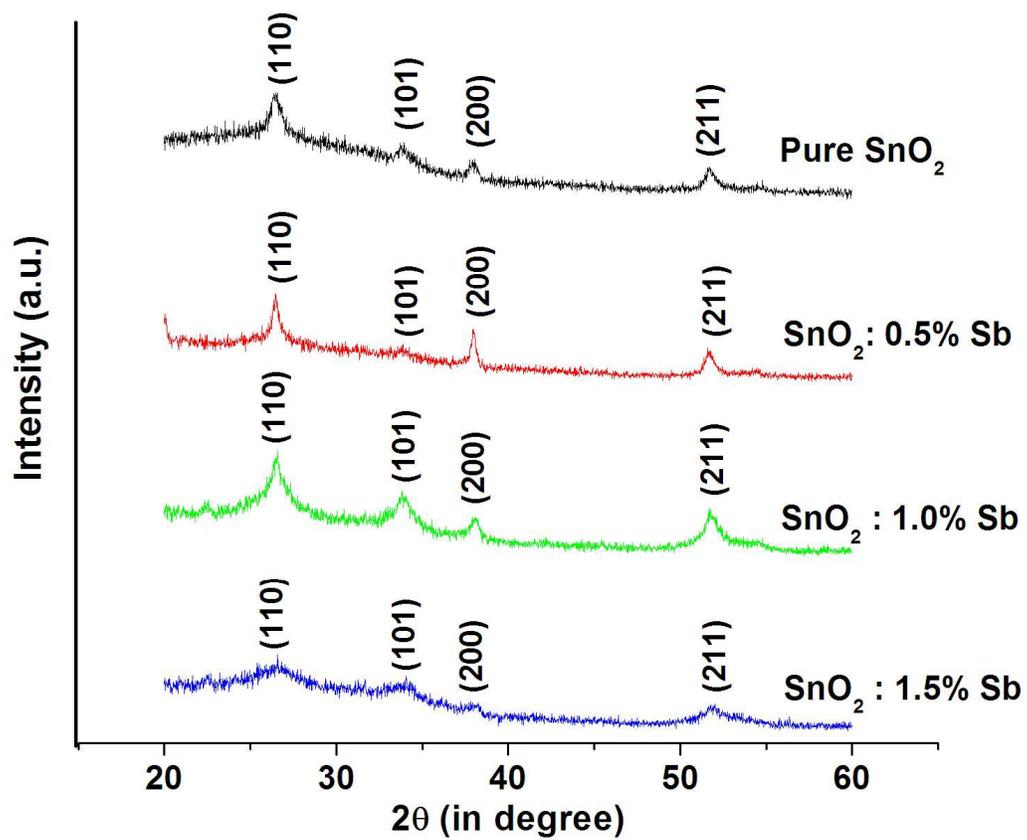

**Figure 1**

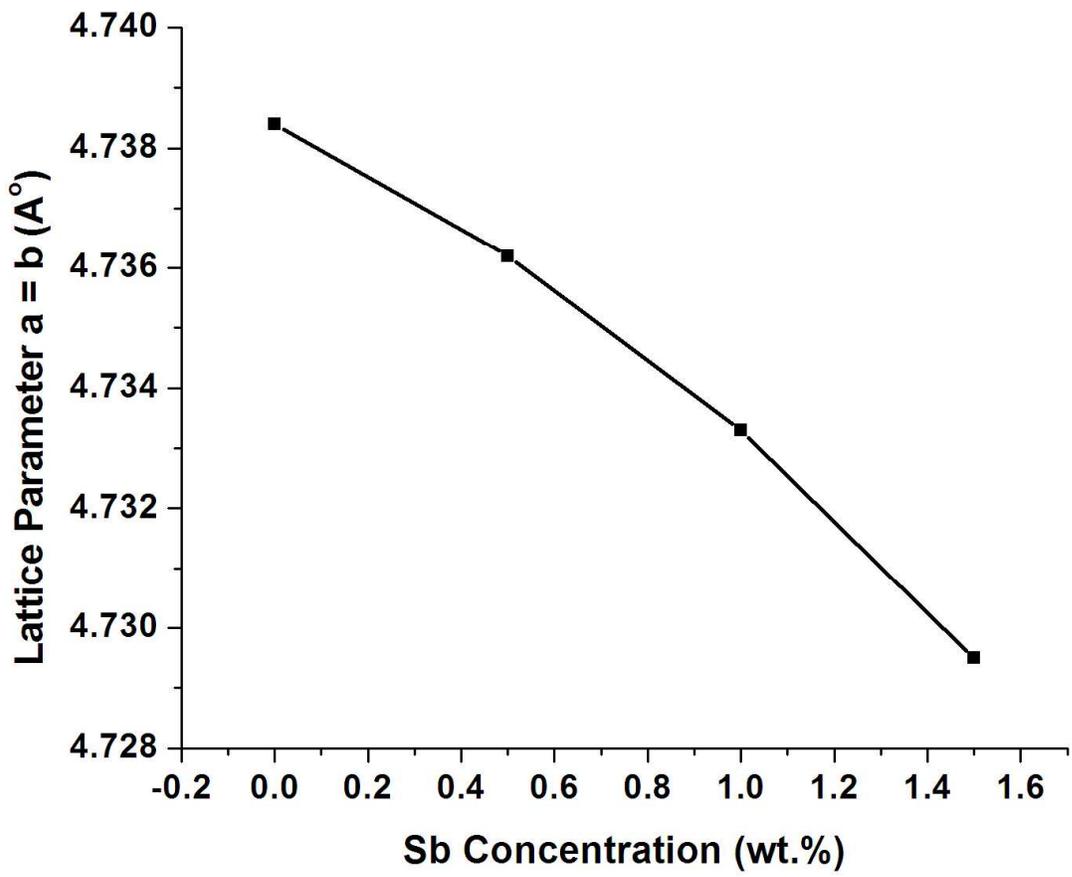

**Figure 2**

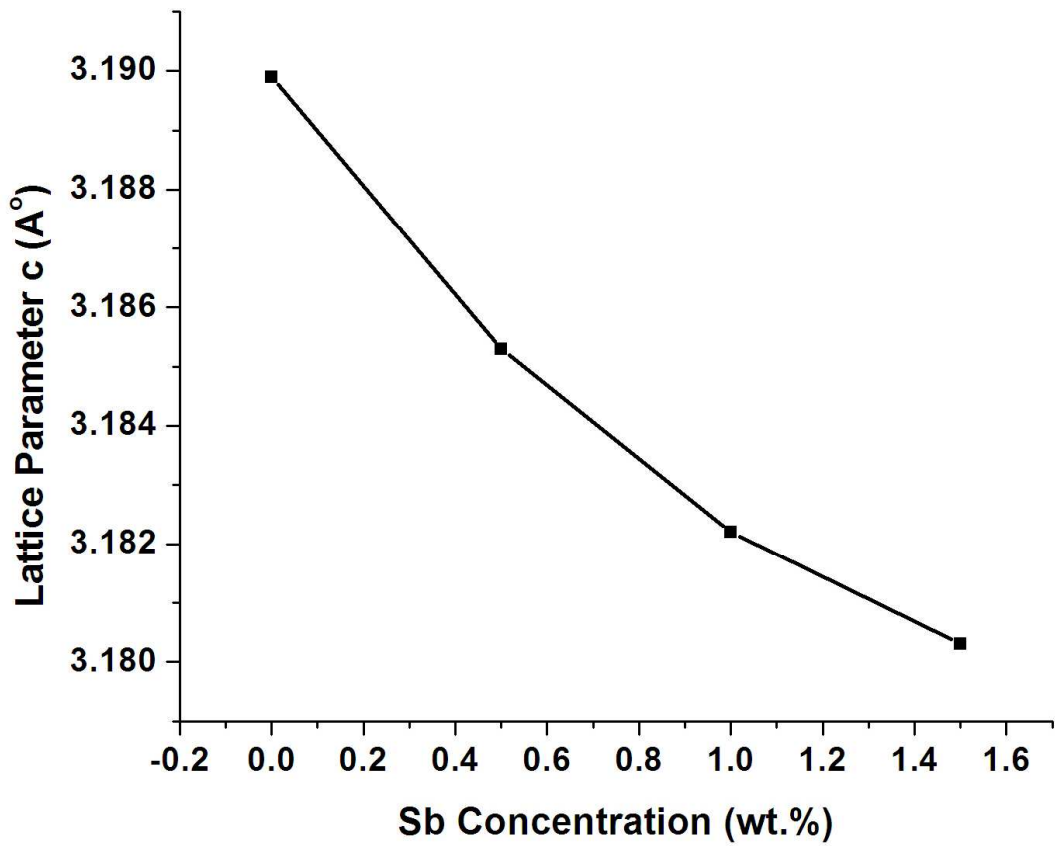

**Figure 3**

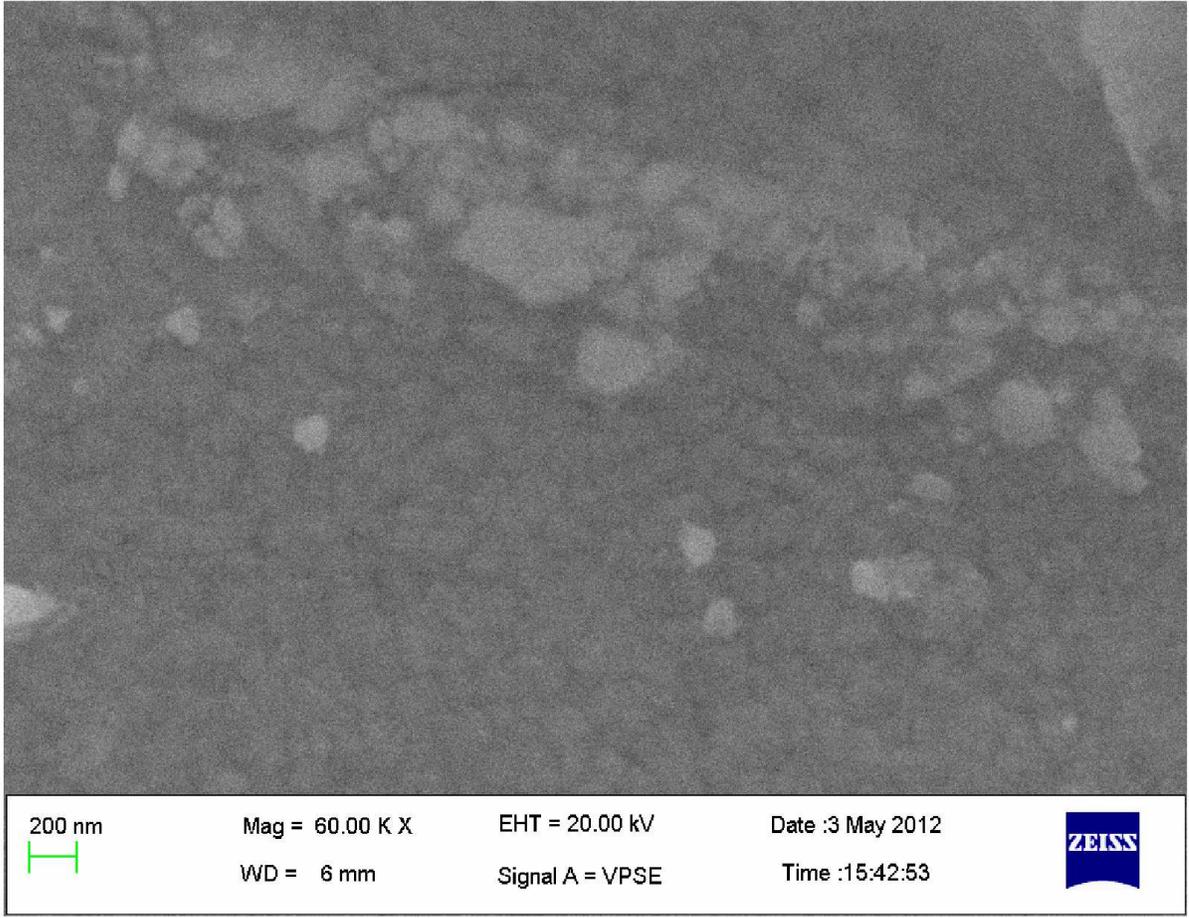

**Figure 4**

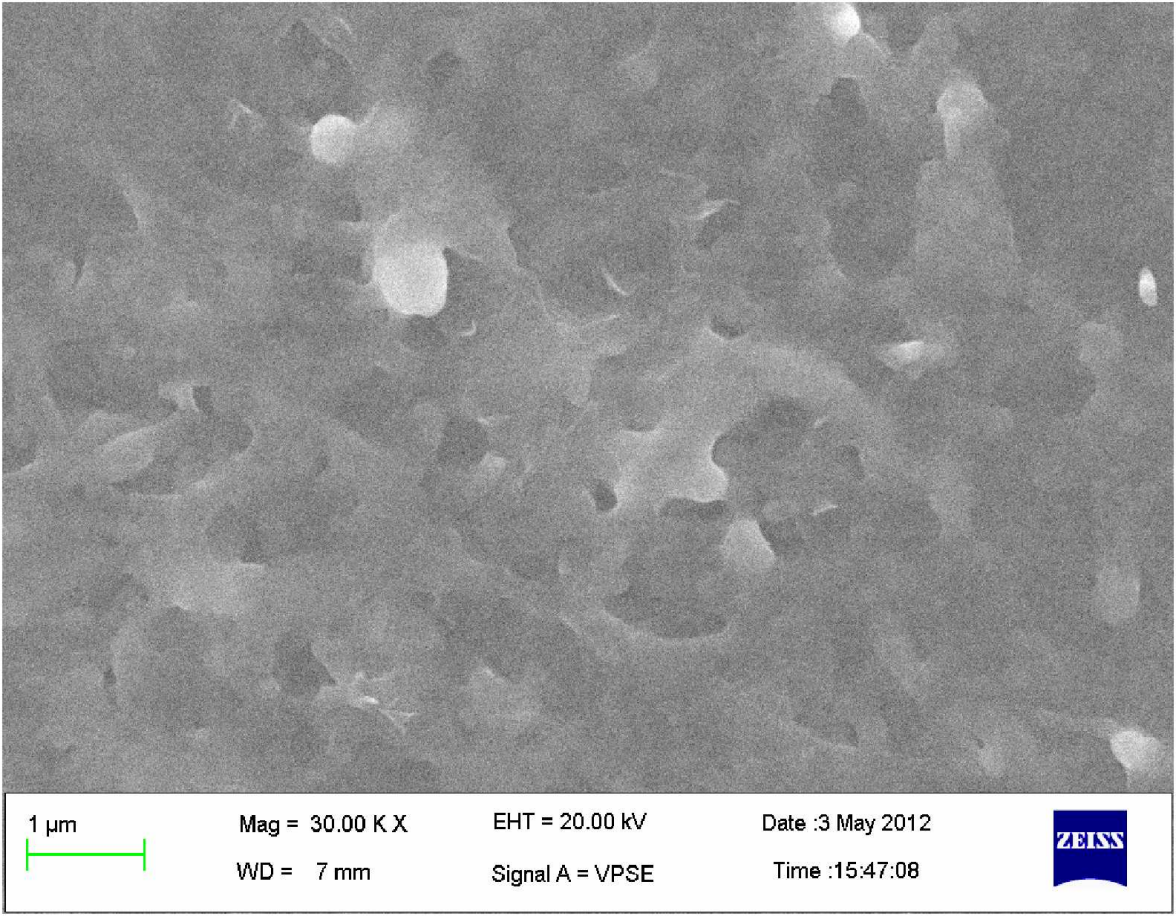

**Figure 5**

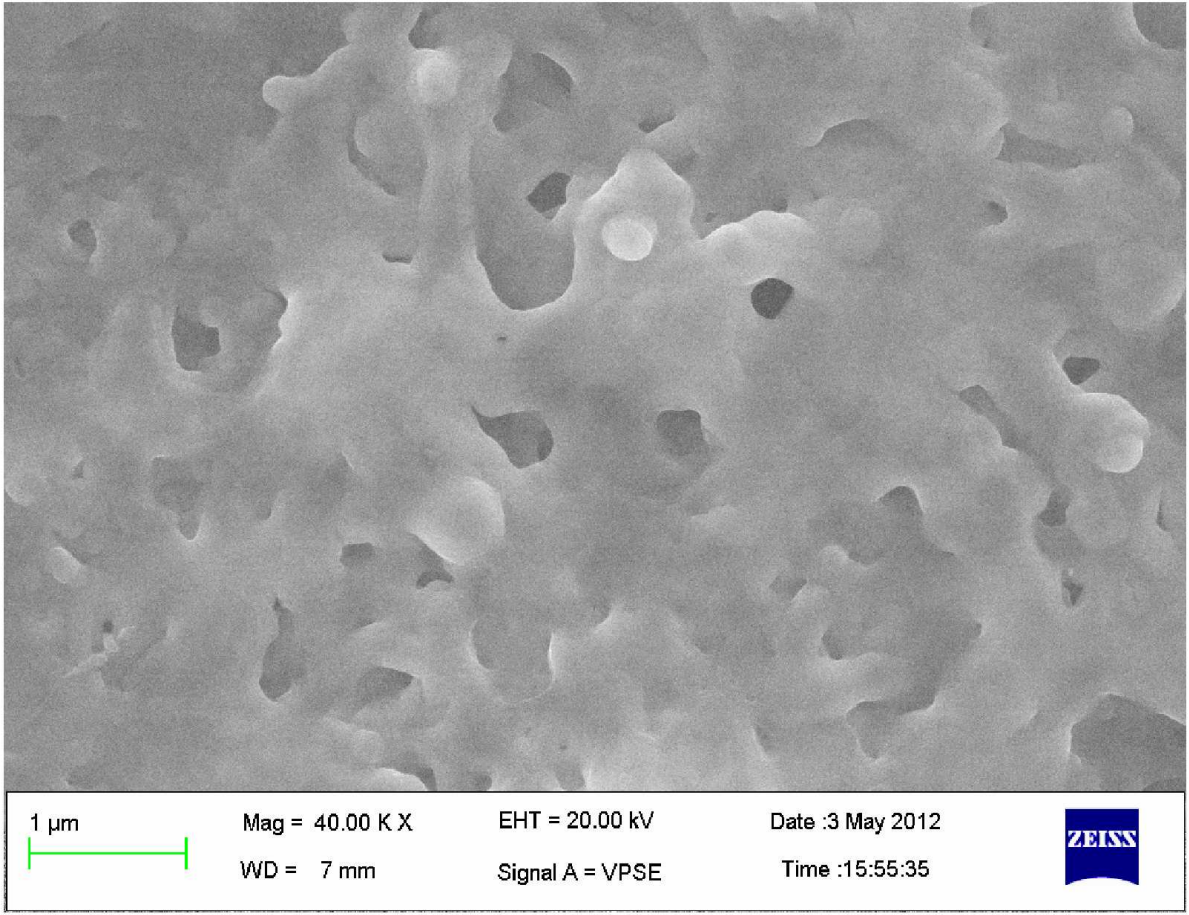

**Figure 6**

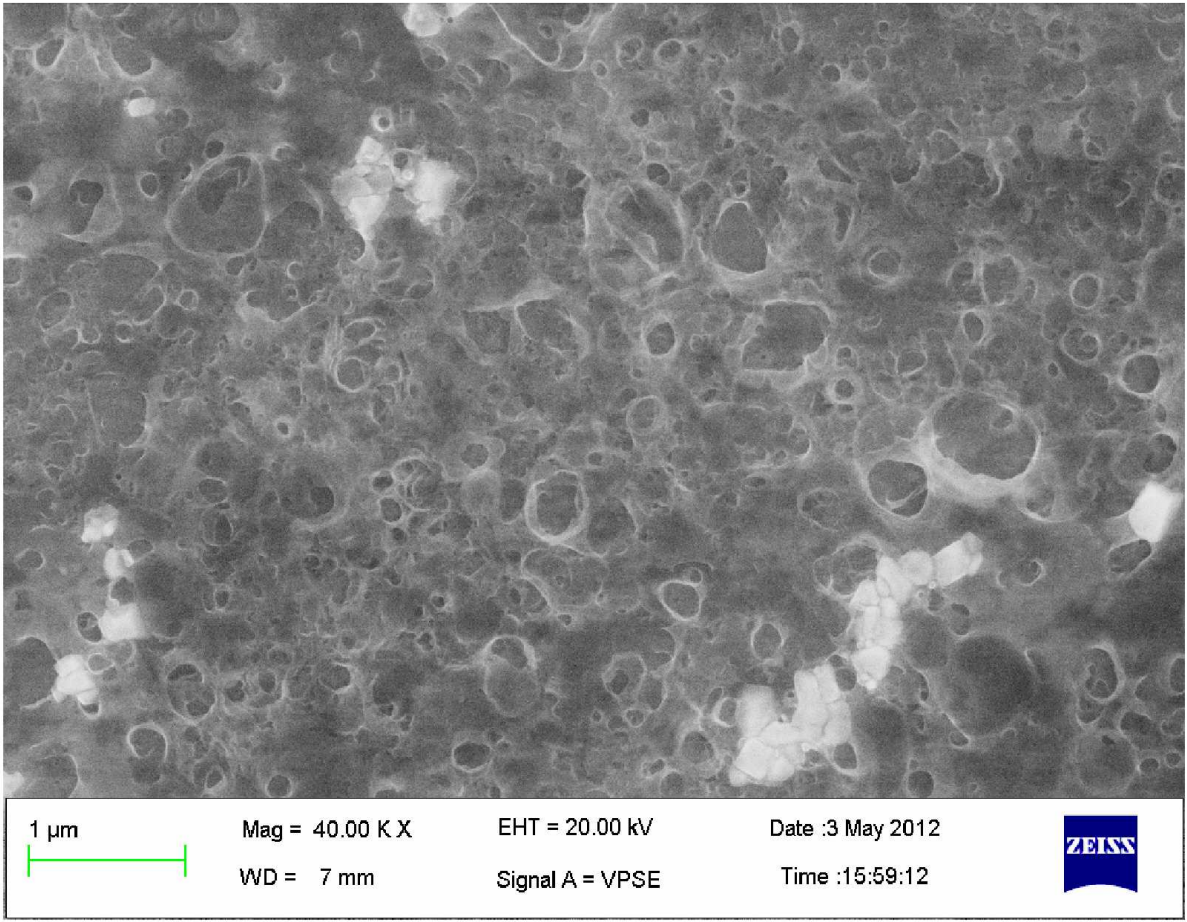

**Figure 7**

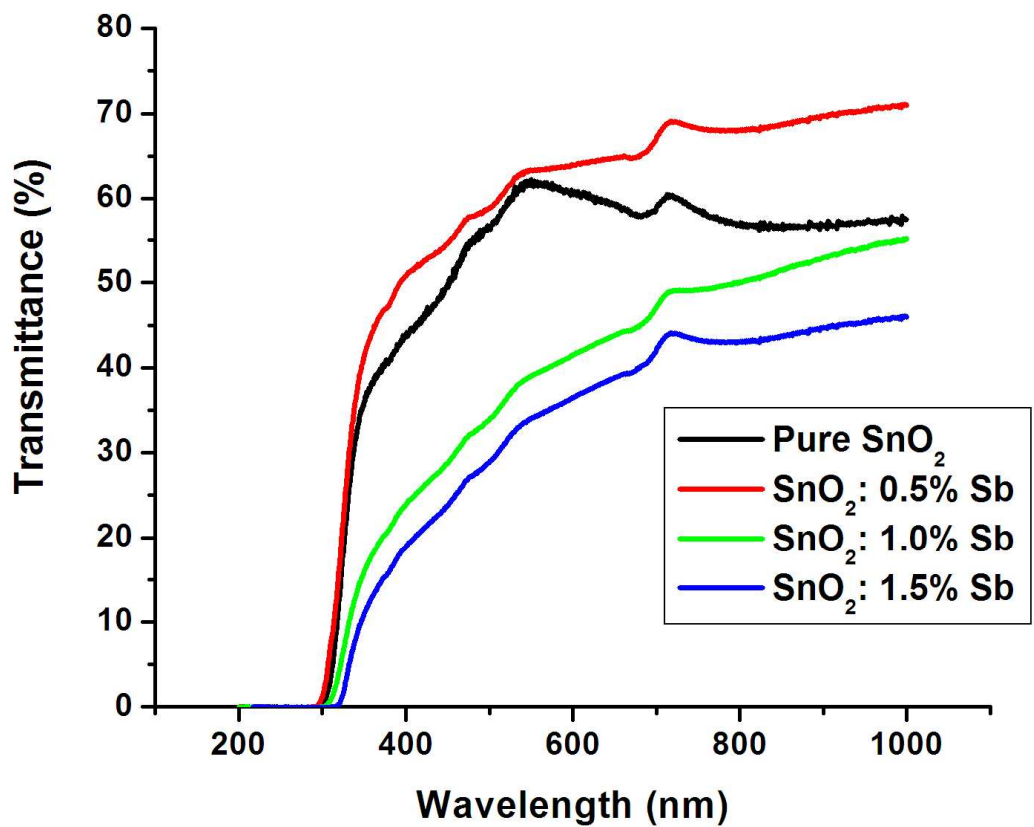

**Figure 8**

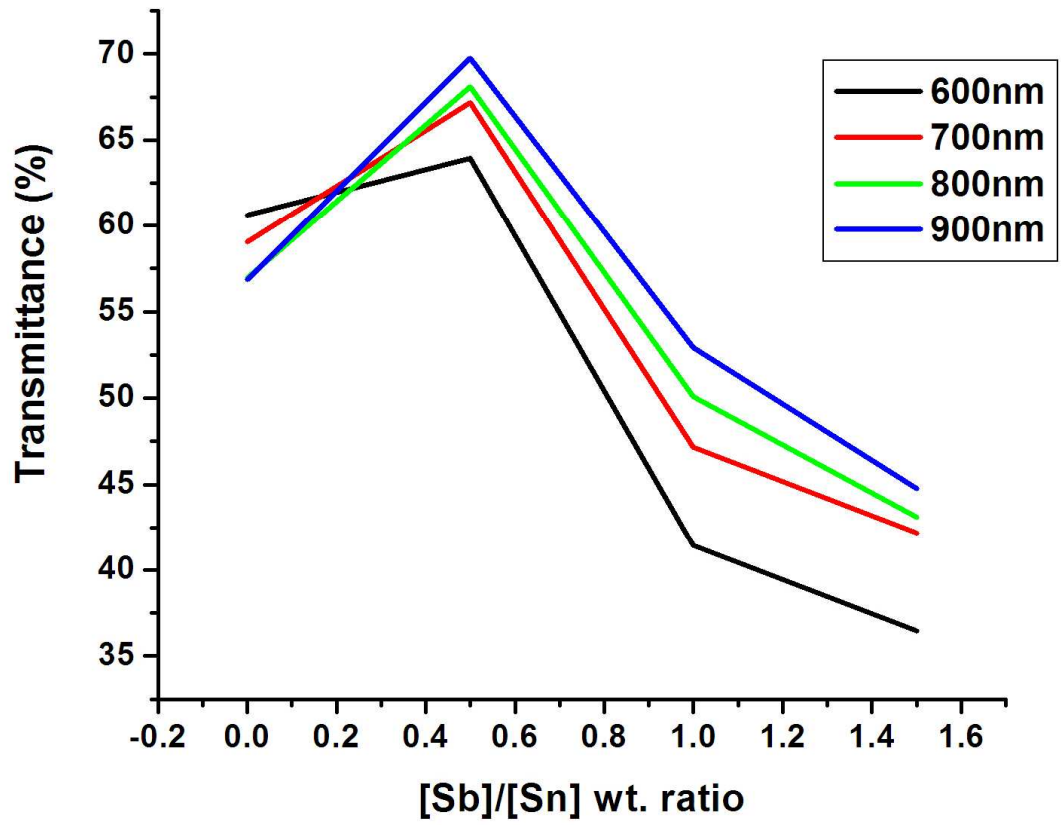

**Figure 9**

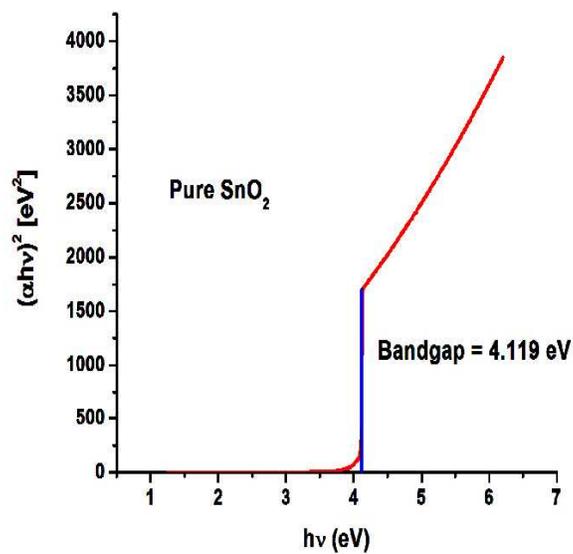
(a)

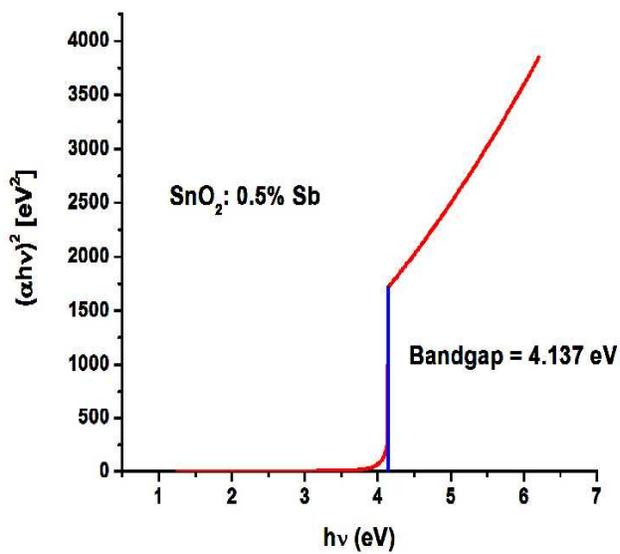
(b)

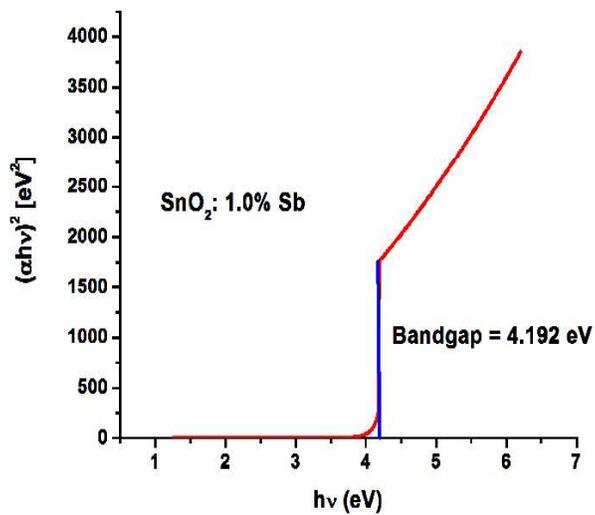
(c)

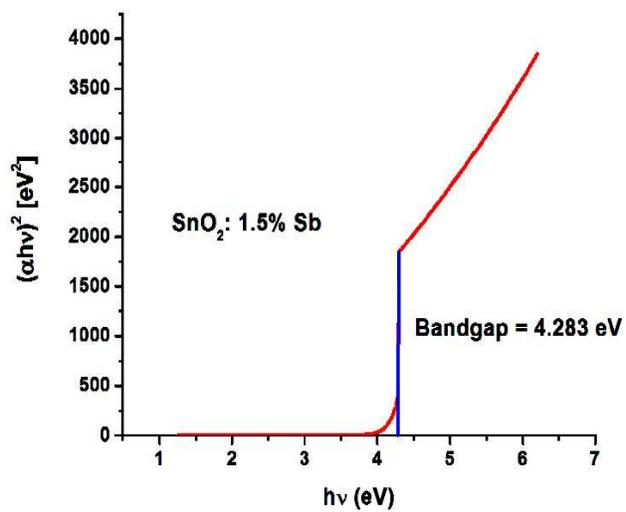
(d)

**Figure 10**

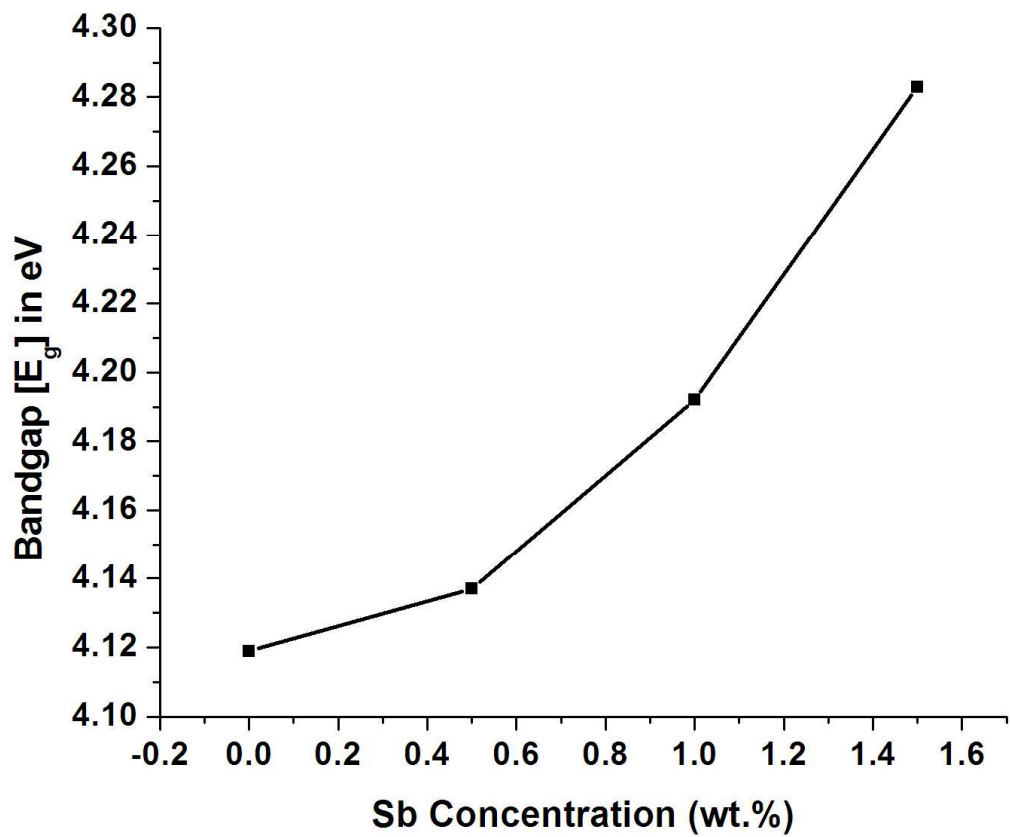

**Figure 11**

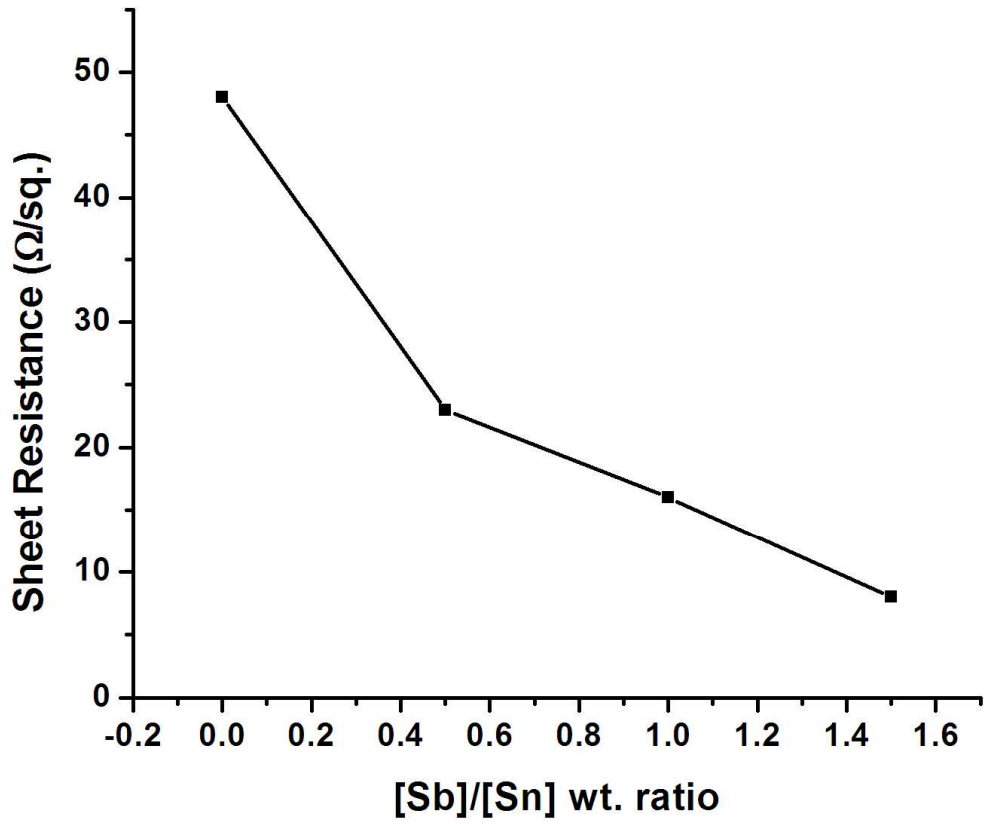

**Figure 12**

# Tables

| Concentration of Sb (wt.%) | Lattice parameters ||
|---|---|---|
| | a = b (Å) | c (Å) |
| [Sb]/[Sn] = 0.0 wt.% | 4.7384 | 3.1899 |
| [Sb]/[Sn] = 0.5 wt.% | 4.7362 | 3.1853 |
| [Sb]/[Sn] = 1.0 wt.% | 4.7333 | 3.1822 |
| [Sb]/[Sn] = 1.5 wt.% | 4.7295 | 3.1803 |

**Table 1**

| Concentration of Sb (wt.%) | Bandgap ($E_g$ (eV)) |
|---|---|
| [Sb]/[Sn] = 0.0 wt.% | 4.119 eV |
| [Sb]/[Sn] = 0.5 wt.% | 4.137 eV |
| [Sb]/[Sn] = 1.0 wt.% | 4.192 eV |
| [Sb]/[Sn] = 1.5 wt.% | 4.283 eV |

**Table 2**